\documentclass[prl,twocolumn,showpacs,amsmath,bibnotes,superscriptaddress]{revtex4}

\usepackage{graphicx}

\newcommand\pictc[5]{\begin{figure}
                       \centerline{
                       \includegraphics[width=#1\columnwidth,height=0.7\textheight,keepaspectratio]{#3}}
                   \protect\caption{\protect\label{fig:#4} #5}
                    \end{figure}            }
\newcommand\pict[4][1]{\pictc{#1}{!tb}{#2}{#3}{#4}}
\newcommand\rpict[1]{\ref{fig:#1}}

\newcommand\leqt[1]{\protect\label{eq:#1}}
\newcommand\reqtn[1]{\ref{eq:#1}}
\newcommand\reqt[1]{(\reqtn{#1})}

\newcounter{Fig}

\begin{document}
\begin{sloppy}

\title{Controlled generation and steering of spatial gap solitons}

\author{Dragomir Neshev}
\author{Andrey A. Sukhorukov}
\affiliation{Nonlinear Physics Group and Centre for Ultra-high
bandwidth Devices for Optical Systems (CUDOS), Research School of
Physical Sciences and Engineering, Australian National University,
Canberra, ACT 0200, Australia}
\homepage{www.rsphysse.anu.edu.au/nonlinear}

\author{Brendan Hanna}
\author{Wieslaw Krolikowski}
\affiliation{Laser Physics Center and Centre for Ultra-high
bandwidth Devices for Optical Systems (CUDOS), Research School of
Physical Sciences and Engineering, Australian National University,
Canberra, ACT 0200, Australia}

\author{Yuri S. Kivshar}
\affiliation{Nonlinear Physics Group and Centre for Ultra-high
bandwidth Devices for Optical Systems (CUDOS),
Research School of Physical Sciences and Engineering,
Australian National University,
Canberra, ACT 0200, Australia}
\homepage{www.rsphysse.anu.edu.au/nonlinear}

\begin{abstract}
We demonstrate the first fully controlled generation of {\em immobile and slow spatial gap solitons} in nonlinear periodic systems with band-gap spectra, and reveal the key features of gap solitons which distinguish them from conventional counterparts, including a dynamical transformation of gap solitons due to {\em nonlinear inter-band coupling}. We also predict theoretically and confirm experimentally the effect of {\em anomalous steering} of gap solitons in optically-induced photonic lattices.
\end{abstract}

\pacs{  42.65.Tg,
    42.65.Jx,
    42.70.Qs 
     }

\maketitle

Nonlinear wave self-action plays an important role in many physical systems ranging from water waves to optical beams and Bose-Einstein condensates (BECs). Nonlinearity can suppress wave spreading leading to the formation of {\em solitary waves} which, due to their robustness, can find many applications, e.g. in long-distance communication systems. Almost two decades ago, it was suggested that the systems with periodically modulated parameters can support a novel type of solitons--{\em gap solitons}, which may exist in band gaps of the linear spectra due to strong wave scattering and coupling between the forward and backward propagating waves~\cite{Voloshchenko:1981-902:ZTF,mills}. It was predicted that gap solitons may exist in different nonlinear periodic structures including fiber Bragg gratings~\cite{deSterke:1994-203:ProgressOptics}, two- and three-dimensional photonic crystals~\cite{Akozbek:1998-2287:PRE}, and BEC in optical lattices~\cite{Zobay:1999-643:PRA}.

According to the theoretical predictions, gap solitons have many unique properties, which distinguish them from conventional solitons~\cite{deSterke:1994-203:ProgressOptics}: (i)~they can form in both self-focusing and self-defocusing nonlinear media; (ii)~soliton velocity and dispersion are nontrivially modified; (iii)~the gap solitons become unstable and undergo dynamical transformations above a critical amplitude due to the inter-band resonances~\cite{Barashenkov:1998-5117:PRL,
Sukhorukov:2003-2345:OL}. For gap solitons observed experimentally in fiber Bragg gratings~\cite{Eggleton:1996-1627:PRL} and waveguide arrays~\cite{Mandelik:2003-53902:PRL}, only some reduction of the soliton velocity was observed. This has prevented an experimental study of dispersion properties of gap solitons and associated band-gap effects, since such effects become evident for slow and immobile solitons only.

In this Letter, we demonstrate the first fully controlled generation of spatial gap solitons where the soliton power and velocity may be independently varied. In particular, we observe experimentally {\em immobile gap solitons}. We demonstrate the gap-soliton steering, reveal their unusual mobility, and observe, for the first time to our knowledge, the gap-soliton dynamics due to instability development. Our experimental results are in excellent agreement with the key theoretical predictions, underlying generic properties of gap solitons with implications to a variety of nonlinear periodic systems beyond the field of optics.

We investigate the gap-soliton formation in photonic lattices created by interfering ordinary polarized laser beams in a biased photorefractive crystal. {\em Discrete solitons} in such lattices were  observed recently~\cite{Fleischer:2003-23902:PRL, Neshev:2003-710:OL}; this allows us to compare two types of solitons emphasizing the intriguing properties of gap solitons.

First, we determine the optimal conditions for gap-soliton generation, using the normalized paraxial equation for the probe beam electric field envelope $E(x,z)$,
\begin{equation} \leqt{nls}
   i \frac{\partial E}{\partial z}
   + D \frac{\partial^2 E}{\partial x^2}
   + {\cal F}( x, |E|^2) E
   = 0 ,
\end{equation}
where $x$ and $z$ are the transverse and propagation coordinates,
normalized to the characteristic values $x_0$ and $z_0$,
respectively, $D = z_0 \lambda / (4 \pi n_0 x_0^2)$ is the beam
diffraction coefficient, $n_0$ is the average medium refractive
index, and $\lambda$ is the vacuum wavelength. The optically
induced change of the refractive index
is~\cite{Fleischer:2003-23902:PRL, Neshev:2003-710:OL}
\begin{equation} \leqt{nonlin}
   {\cal F}( x, |E|^2) = - \gamma (I_b + I_p(x) +
  |E|^2)^{-1},
\end{equation}
where $I_b$ is the constant dark irradiance, $I_p(x)$ is the
two-beam interference pattern which induces the periodic lattice
with a period $d$,
\begin{equation} \leqt{periodic}
   I_p(x)= I_g \cos^2(\pi x / d),
\end{equation}
and $\gamma$ is a nonlinear coefficient proportional to the
applied DC field. In order to match our experimental conditions,
we use below the following parameters for the theory and numerical
calculations: $\lambda = 0.532~\mu$m, $n_0 = 2.4$, $x_0 = 1~\mu$m,
$z_0 = 1$mm, $d = 22.2$, $I_b =1$, $I_g = 1$, $\gamma = 5.31$, the
crystal length is $L = 15$~mm. We note that the model
Eq.~\reqt{nls} is very general and appears in other areas of physics. It can
describe, in particular, the BEC dynamics in optical lattices where the matter-wave gap solitons are expected to have similar properties to optical counterparts.

Existence of spatial bright solitons is closely linked to the
structure of the linear wave spectrum. In periodic lattices, the
spectrum is composed of bands corresponding to the propagating
{\em Floquet-Bloch modes}, which are separated by gaps where the
wave propagation is forbidden. The Floquet-Bloch modes are
solutions of linearized Eq.~\reqt{nls} in the form
$E_{\kappa,n}(x,z) = \psi_{\kappa,n}(x) \exp(i \kappa x/d + i
\beta_{\kappa,n} z)$, where $\beta_{\kappa,n}$ and $\kappa$ are
the Bloch-wave propagation constant and wavenumber, respectively,
and the index $n = 1,2,\ldots$ marks the order of the transmission
band. The Bloch wave $\psi_{\kappa,n}(x)$ has the periodicity of
the photonic structure, $\psi_{\kappa,n}(x+d) \equiv
\psi_{\kappa,n}(x)$, and this condition defines the dispersion
relation $\beta_{\kappa,n}$, as sketched in
Fig.~\rpict{dispers}(a). The open regions in this plot mark band
gaps where ${\rm Im} \kappa \ne 0$: the top one exists due to
total internal reflection and extends to $\beta \rightarrow
+\infty$, whereas lower gaps have a finite width and appear due to
Bragg scattering from the periodic structure.

\pict{fig01.eps}{dispers}{ (a) Dispersion of Bloch waves in an
optically-induced lattice; the spectrum bands are shaded. (b)
Bloch-wave profiles (solid) and leading-order Fourier components
(dashed) superimposed on top of the normalized refracted index
profile of the periodic lattice (shown with shading) for different
gap edges indicated by the arrows. }

In order to underline the essential physics of the soliton
formation in periodic lattices, first we consider the
small-amplitude limit. We seek solutions of Eq.~\reqt{nls} near
the band edge
in the form of modulated Bloch waves, $E = \varphi(x,z)
\psi_{\kappa,n} \exp(i \beta_{\kappa,n} z + i \kappa x / d)$, and
derive the nonlinear Schr\"odinger equations for the slowly
varying Bloch-wave envelope~\cite{Sipe:1988-132:OL},
\begin{eqnarray} \leqt{nlsc}
  i \frac{\partial \varphi}{\partial z}
  + i V_g \frac{\partial \varphi}{\partial x}
  + \frac{\eta}{2} \frac{\partial^2 \varphi}{\partial x^2}
  + \Gamma |\varphi|^2 \varphi
  = 0.
\end{eqnarray}
Here $V_g = -d \left. \partial \beta / \partial \kappa
\right|_{\beta_{\kappa,n}}$ is the group velocity, $\eta = - d^2
\left. \partial^2 \beta / \partial \kappa^2
\right|_{\beta_{\kappa,n}}$ is the effective diffraction
coefficient, and $\Gamma = \int_0^d \left[{\cal F}( x,
|\psi_{\kappa,n}|^2)-{\cal F}( x, 0)\right] |\psi_{\kappa,n}|^2
dx$ is the effective nonlinear coefficient, where we assume the
normalization $\int_0^d |\psi_{\kappa,n}|^2 d x \equiv 1$.
Equation~\reqt{nlsc} possesses localized solutions for bright
solitons, $\varphi = \sqrt{2 \rho/\Gamma} {\rm sech}\left[  (x -
V_g z) \sqrt{\rho/\eta} \right] \exp(i \rho z)$, provided
$\eta\Gamma >0$ and $\rho$ is a free parameter. Thus, for
self-focusing medium nonlinearity ($\Gamma>0$), the gap solitons
appear near the lower-gap edges of the bandgap spectrum.

Next, we find soliton solutions of the full model
Eqs.~\reqt{nls}-\reqt{periodic}, shown in Fig.~\rpict{power}, and
identify a number of important differences between the
conventional and gap solitons which appear when the propagation
constant is shifted deep inside the gap, where the envelope
approximation~\reqt{nlsc} is no longer valid. Specifically, the
power of gap solitons is bounded from above
[Fig.~\rpict{power}(top)], and their width is bounded from below
[Fig.~\rpict{power}(middle)]. These limitations appear because the
Bragg-reflection gap always has a finite width, in contrast to the
semi-infinite gap where conventional (or discrete) solitons are
localized.

\pict{fig02.eps}{power}{ Numerical results for a gap soliton family:
Power (top) and width (middle) vs. the propagation constant.
Bottom: soliton profiles (solid) corresponding to the marked
points (a) and (b) in the upper plots; shadings mark the grating
minima. Arrows illustrate schematically the direction of the input
beams, which interference pattern (shown with dashed line)
approximates the soliton profile. }

Controlled experimental excitation of spatial gap solitons can be
realized if the modulated Bloch-wave profile is properly matched
at the input. Since the Bloch waves are periodic, they can be
decomposed into the Fourier series, $\psi_{\kappa,n} = \sum_m
C_n(\kappa+2 \pi m) \exp[i x (\kappa+2 \pi m) /d)$. In
Fig.~\rpict{dispers}(b) we show the characteristic profiles of the
Bloch waves, and also plot the contribution from the leading-order
Fourier components (dashed lines). We find that in the leading
order  $\psi_{0,1} = C_1(0) + \ldots$, therefore lattice solitons
in the semi-infinite total internal reflection gap can be
generated by {\em a single incident beam}, as was realized in
earlier experiments for arrays of weakly coupled optical
waveguides. On the contrary, the Bloch waves in the
Bragg-reflection gap are composed of {\em the counter-propagating
waves}, e.g. $\psi_{\pi,n} = C_{n}(\pi) \exp(i \pi x/d) +
C_{n}(-\pi) \exp(-i \pi x/d) + \ldots$ for $n=1,2$. Therefore,
spatial gap solitons can be generated by using {\em two Gaussian
beams} which are tuned to the Bragg resonance and have opposite
inclination angles, as was originally suggested in
Ref.~\cite{Feng:1993-1302:OL} and developed further in
Ref.~\cite{Sukhorukov:2003-2345:OL}. Therefore, we consider the
input electric field in the form,
\begin{equation} \leqt{interfInput}
 E_0(x) = \sqrt{I_0} e^{ - (x-x_e)^2 / w^2 }
            \cos\left[ \pi (x - x_s) / d \right],
\end{equation}
where the exponential term approximates the gap soliton envelope,
$w$ being the width of the input beams, and the interference term
approximates the Bloch-wave profile, with the shift $x_s$
depending on the relative phase difference between the two beams,
and we choose $x_e = d/2$ and $w = 55$.
When the interference maxima are at the minima of the refractive
index profile, the Bloch mode is excited at the lower edge of the
Bragg-reflection gap, and the input pattern can match very closely
the gap-soliton profile, as shown in examples of
Figs.~\rpict{power}(a,b).

We investigated numerically the dynamics of the two-beam mutual
focusing and the gap-soliton generation by simulating the model
Eq.~\reqt{nls} with the input condition~\reqt{interfInput} chosen
to match the Bloch-wave profile at the lower gap edge. The beams
diffract at a low input power [Fig.~\rpict{focus}(a)], whereas an
immobile gap soliton forms when the input power is increased
[Fig.~\rpict{focus}(b)]. We note that the required power depends
on the input beam width, as follows from Fig.~\rpict{power}, and
the minimum soliton width defines a fundamental limitation of the
degree of two-beam mutual focusing. Indeed, as the power is increased,
an instability rapidly develops through a resonant excitation of the
first band, and subsequent formation of a quasi-periodic breather
[Fig.~\rpict{focus}(c)]. Similar breathing states were recently observed
in waveguide arrays~\cite{Mandelik:2003-253902:PRL}.

\pict{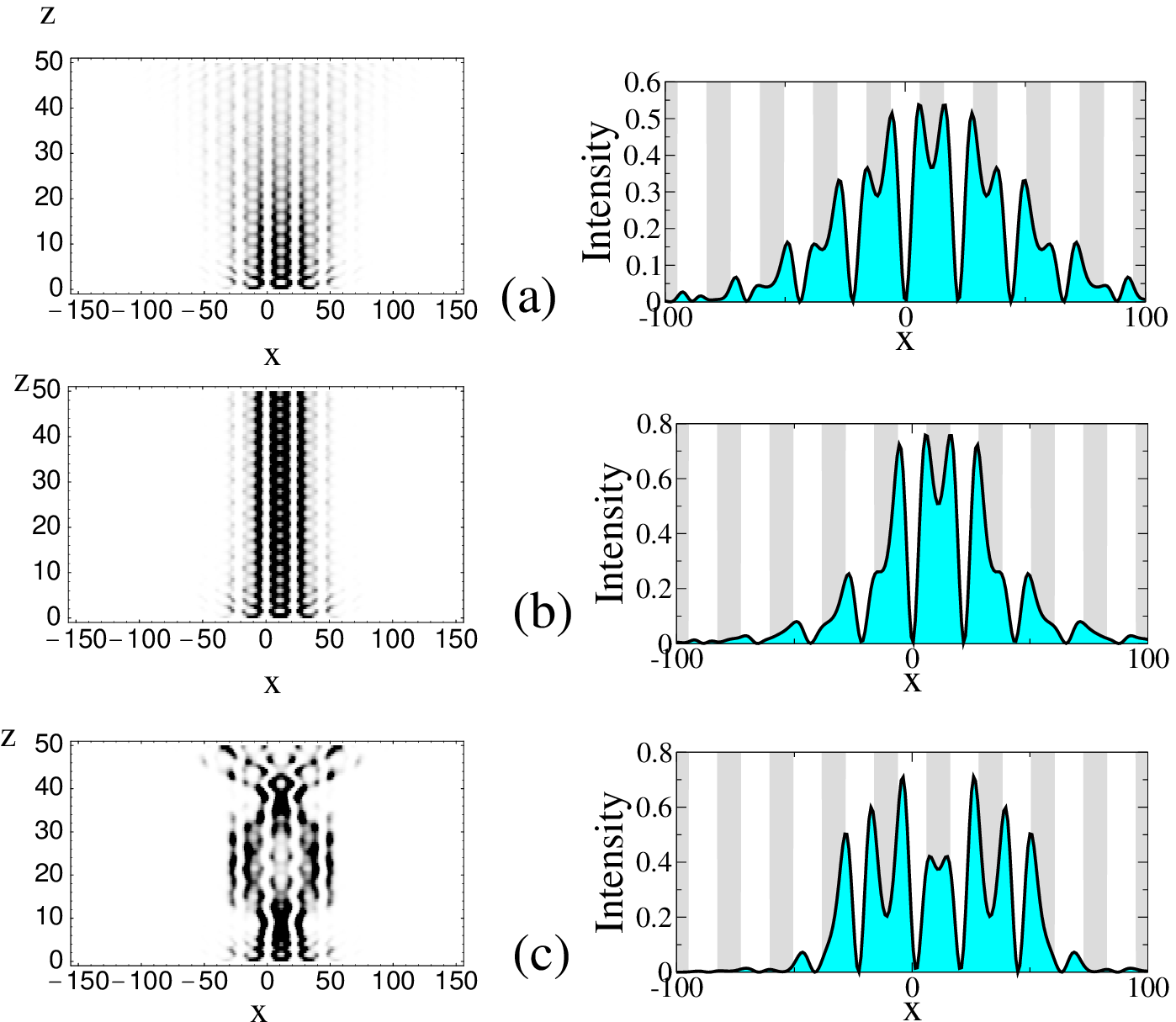}{focus}{Numerical results. Dynamics of the Bloch waves
excited through two-beam interference: (a)~linear diffraction at
low power ($I_0 \simeq 0$), (b)~excitation of a gap soliton in the
nonlinear regime ($I_0=0.048$), (c)~beam breakup and the formation
of a quasi-periodic breathing state at higher powers ($I_0=0.28$).
Left: variation of intensity along the propagation direction; Right:
beam profiles at the crystal output ($z=15$mm) normalized to $I_0$. }

Our experiments were performed in a 15mm-long Strontium Barium
Niobate (SBN:60) crystal externally biased along the crystalline
$c$-axis. The experimental setup is similar to that discussed in
Ref.~\cite{Neshev:2003-710:OL} with the difference that the
extraordinary polarized probe beam was split into two parts which were
later focused by cylindrical lens and made to overlap at the input
face of the crystal. The angle between these two beams was set
to twice the Bragg angle, such that the periodicity of the
interference pattern is equal to that of the lattice (22~$\mu$m).
In our case, such value of the periodicity allows for a relatively
wide gap in the transmission spectrum, as shown in Fig.~\rpict{dispers}(a).
The relative phase between the probe beams was tuned such that a
symmetric interference pattern is obtained, as shown in
Fig.~\rpict{gapExper}(top). The relative
position between this pattern and the lattice could also be
controlled, by changing the relative phase between the two lattice
forming beams~\cite{Neshev:2003-710:OL}. The input width of the
overlapping probe beams is $w=55\mu$m (65$\mu$m FWHM), while they
are fully separated at the output for zero bias field. When
electric field of 5000 V/cm is applied to the crystal, the index
grating (optical lattice) is formed and the probe beams excite
Bloch waves in the first or the second band, depending on the
relative lattice position.

\pict{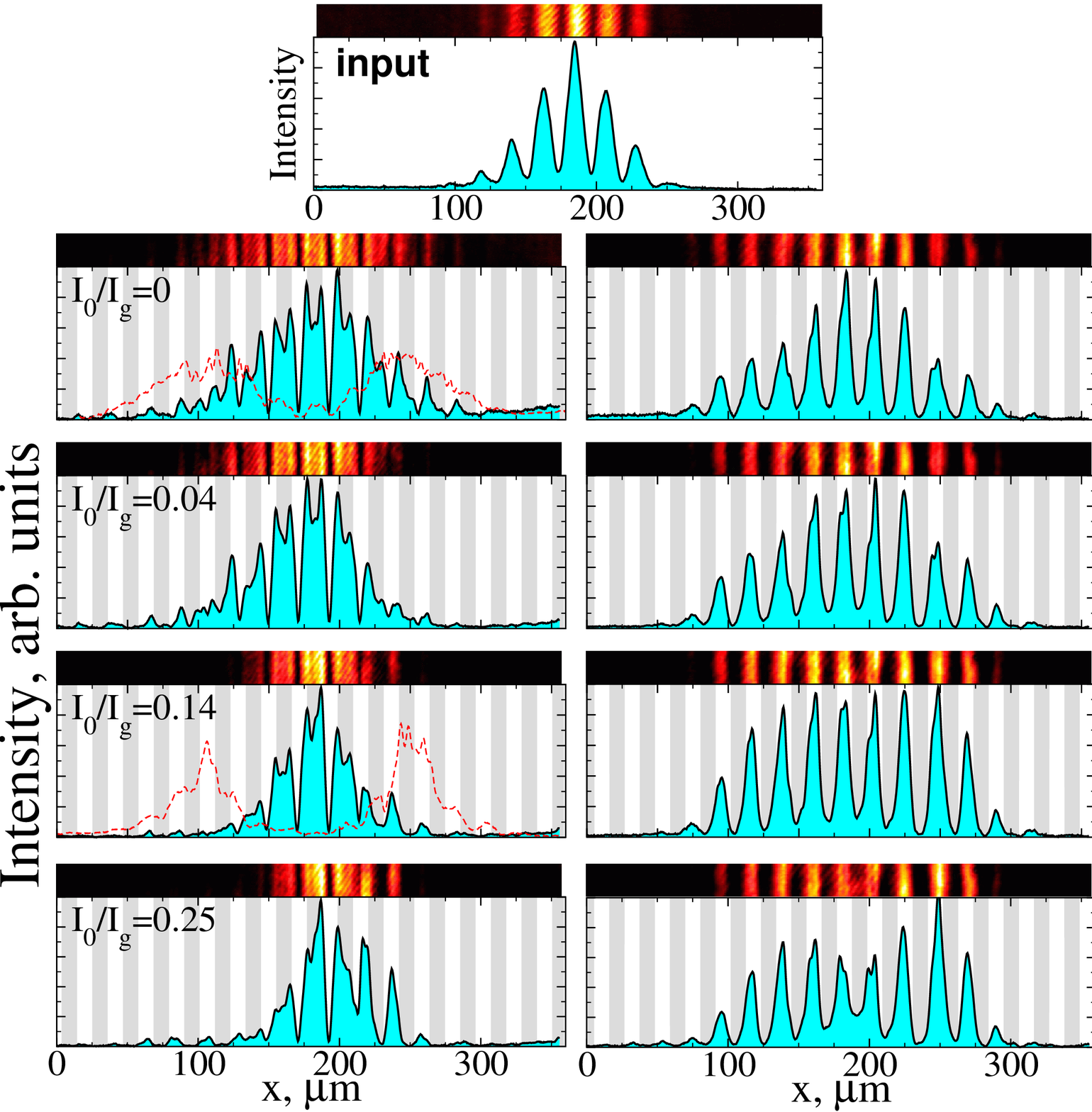}{gapExper}{Experimental results for two-beam
interaction showing the intensity profiles at the crystal input (top)
and output for varying beam power: Left: mutual focusing and gap soliton
formation when interference maxima are aligned with the lattice
minima; Right: self-defocusing when interference maxima of the
input beams are at the lattice maxima. Dashed curves represent the
beam profiles at the indicated beam intensity when the grating is
erased. }

\pict{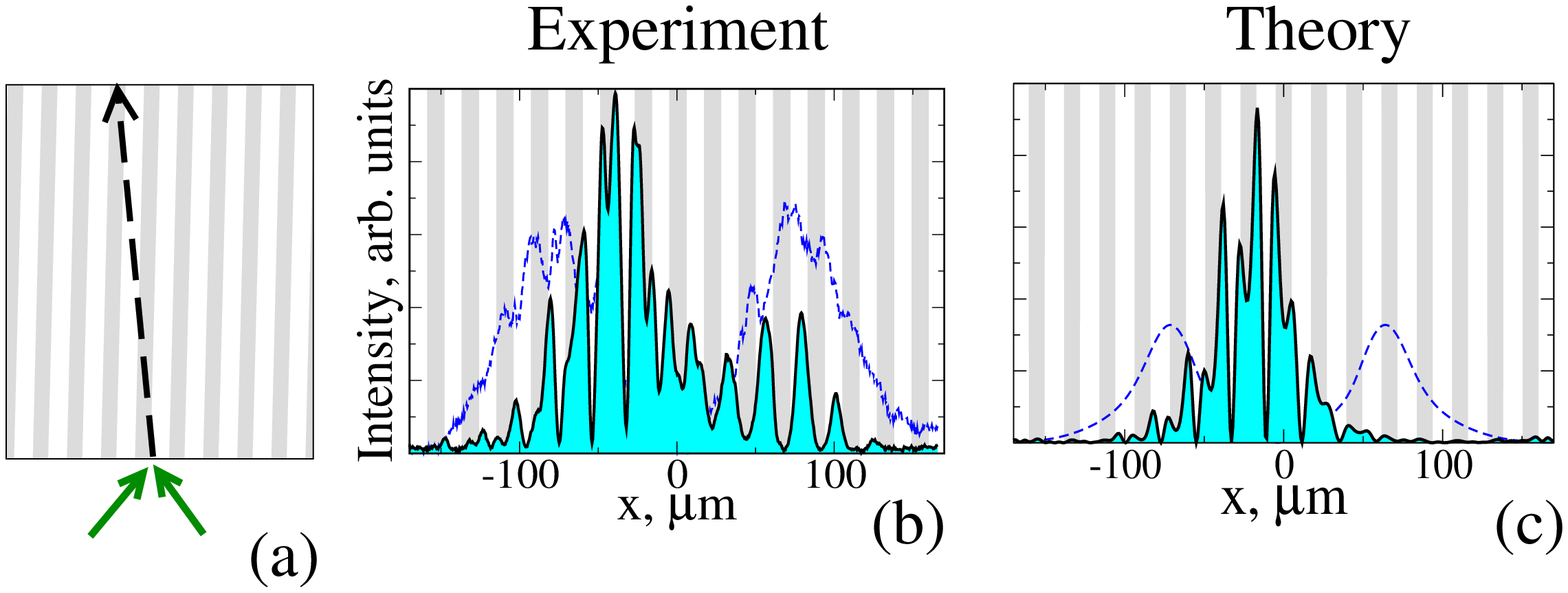}{steering}{ (a) Schematic demonstration of
anomalous gap-soliton steering induced by a tilt of the lattice;
dashed line shows the propagation direction of a gap soliton;
solid lines indicate the directions of the input beams. (b,c)
Output soliton profile for a lattice tilt in the direction of
larger $x$ by 20\% of the Bragg angle with respect to normal;
dashed lines show the beam profiles when the lattice is absent.}

The experimental results shown in Fig.~\rpict{gapExper} represent
remarkably all theoretical predictions. First, we align the
interference maxima of the input beams with the minima of the
induced lattice at the input face of the crystal and record the
beam profiles at back face for several input powers [see
Fig.~\rpict{gapExper}(left column)]. The output intensity is
exactly zero at the maxima of the index grating. The intensity
maxima are out-of-phase, as confirmed with interferometric
measurements, and posses a double peak structure located at the
minima of the grating. At low powers, the output intensity pattern
is broad and corresponds to the Bloch waves at the lower gap edge.
The beam is exactly centered between two unperturbed output beams
[dashed curve in Fig.~\rpict{gapExper}(left column, top plot)]
measured for zero voltage. At higher intensity ($I_0/I_g=0.04$),
we observe mutual focusing, and self-trapping into a spatial gap
soliton is observed at $I_0/I_g=0.14$ [see
Fig.~\rpict{gapExper}(left column)]. The generated gap soliton has
zero transverse velocity, and is centered between the two output
beams which separate if the grating is erased (dashed curve). As
predicted theoretically, the effect of the mutual focusing is
limited and at higher intensities ($I_0/I_g=0.25$) the beam
disintegrates [see Fig.~\rpict{gapExper}(left column, bottom
plot)] while its profile becomes asymmetric due to the diffusion
contribution to the photorefractive nonlinearity. On the other
hand, when we align the interference maxima of the input beams to
the lattice maxima, the excited Bloch wave corresponds to the
upper gap edge [see Fig.~\rpict{dispers}] and they experience
anomalous diffraction ($\eta<0$) leading to self-defocusing as the
power is increased [Fig.~\rpict{gapExper}(right)].

Finally, we study mobility of spatial gap solitons and a
possibility to vary their transverse velocity. Experimentally, the
soliton motion is induced by tilting the lattice by 20\% of the
Bragg angle thus introducing a lateral shift of the induced
waveguides by 16$\mu$m at the output [the lattice is shifted to
the right in Fig.~\rpict{steering}(a)]. 
Results of the experimental observations and the
numerical simulations are presented side-by-side in
Fig.~\rpict{steering}(b,c) show that the generated
gap solitons {\em move to the left} when the grating is {\em tiled
to the right}. In experiment, on the right-hand side of the gap
soliton, a small contribution from the first band is observed,
which appears due to asymmetry in the initial excitation and
inhomogeneities of the lattice. This {\em anomalous steering}
behavior occurs because the spatial group-velocity dispersion for
gap solitons is almost three times larger compared to a
homogeneous crystal. This is in sharp contrast with conventional
lattice (or discrete) solitons, that tend to propagate along the
lattice~\cite{Morandotti:1999-2726:PRL}.

In conclusion, we have demonstrated experimentally the first fully
controlled generation of spatial gap solitons in optically-induced
periodic photonic lattices and observed novel effects such as
anomalous steering of gap solitons and the limitation of the
two-beam mutual focusing through inter-band coupling. We believe
our results can be useful for the study of nonlinear effects in
photonic crystals and nonlinear dynamics of the Bose-Einstein
condensates in optical lattices.

\end{sloppy}
\end{document}